# RAMAN SPECTROSCOPY of the EUTECTIC in the MIXED CRYSTAL of the PARA-DIBROMOBENZENE with the PARA-CHLORONITROBENZENE

**M. A. Korshunov**[1]

Kirenskii Institute of Physics, Siberian Division, Russian Academy of Sciences, Krasnoyarsk, 660036 Russia

**Abstract** – The mixed crystal of a para-dibromobenzene with a para-chloronitrobenzene is investigated at concentration of components from 0% up to 60% of a para-chloronitrobenzene by the method of Low-Frequency Raman spectroscopy. It is shown, that in range of concentrations from 25% up to 50% of a para-chloronitrobenzene the spectrum of the mixed crystal would consist of the sum of spectrums α and β phases which relation of intensities depends on concentration of components. It is also found, that the single crystal in this range has rod frame.

The typical example of the molecular mixed crystal having an eutectic is the solid solution of a para-dibromobenzene with a para-chloronitrobenzene [1]. This crystal has two points of an eutectic at concentration of a para-chloronitrobenzene of 35.5% and 78.0%.

Under X-ray diffraction data [1,2] in range of concentrations from 0.0% up to 28.0% of a para-chloronitrobenzene the α phase with frame of a pure para-dibromobenzene is located. In range from 44.0% up to 74.0% the solid solution crystallizes in one of space groups C2/m, C2 or Cm (β phase). Above 78.0% there is a γ phase with frame of a para-chloronitrobenzene. All three phases have two molecules in a unit cell.

The range from 28.0% up to 44.0% of a para-chloronitrobenzene is insufficiently studied.

We brought up single crystals of solid solutions of a para-dibromobenzene with a para-chloronitrobenzene for lines of concentrations of components on Bridgman's method [3]. The single crystal was cut on a series of tablets, which were investigated on homogeneity. At study of sections of single crystals through the polarizing microscope at concentration of components from 25.0% up to 50.0% of a para-chloronitrobenzene appearance of rod frame is observed [4]. In figure 1 sections along grow of a crystal (1a) and perpendicular to direction of grows (1b) are shown. An appearance of rod frame obviously speaks about appearance in the mixed crystal α and β phases. That it to confirm Raman spectrums of investigated samples were obtained. Concentration of components in mix-crystals was determined on a relation of intensities and shift of frequencies of the intramolecular oscillations. Spectrums of oscillations of the lattice were obtained at concentration of a para-chloronitrobenzene from 0.0% up to 60.0%.

---
[1] E-mail: makorshunov@mail.ru

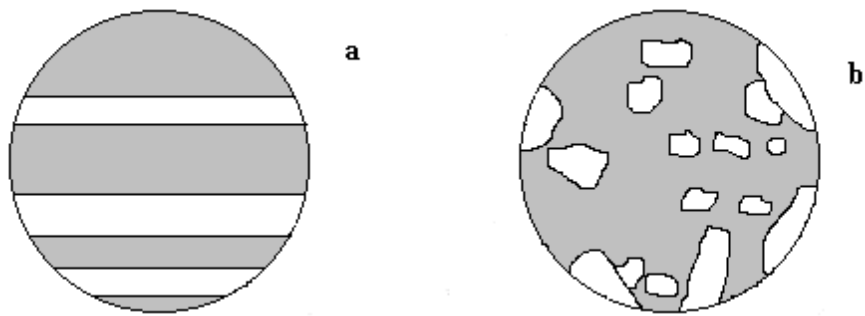

Fig. 1. Microphotographs of sections along grow of a crystal (a) and perpendicular to direction of grows (b).

Polarizing researches of spectrums of solid solutions of a para-dibromobenzene with a para-chloronitrobenzene in a phase at concentration of a para-chloronitrobenzene of 20.0% would show, that in a spectrum are present six intensive lines and a series of padding lines of small intensity. Intensive lines with frequencies 19.5, 36.5, 37.2, 38.7, 90.0 and 93.0 cm$^{-1}$ are connected to rotary swinging of molecules. Additional lines of small intensity 28.0, 32.0 and 58.0 cm$^{-1}$ apparently are connected to translational oscillations. Appearance of lines with frequencies 72.0 and 83.0 cm$^{-1}$ is stipulated by presence of imperfections (vacancy, orientation and space randomness of molecules of components on the lattice of the mixed crystal). Padding lines connected to presence β phases it is not found. In sections of a single crystal of a solid solution at this concentration rod frame is also not found.

In figure 2 low – frequency spectrums of powders of the investigated mix-crystals at a various content of components (a - 17.0%, b - 34.5%, c - 41.0% and d - 50.0% of a para-chloronitrobenzene). The line near 20.0 cm$^{-1}$ is the narrowest and is far from other lines. Apparently from figure 2, intensity of this line with increase of concentration of a para-chloronitrobenzene decreases. Thus intensity of a line in range 85.0 cm$^{-1}$ is increased, and lines 97.0 cm$^{-1}$ decreases. The first line falls into a spectrum of β phases, and the second to a spectrum of α phase. Both lines mostly clear visible at concentration of 41.0% of a para-chloronitrobenzene (fig. 2c).

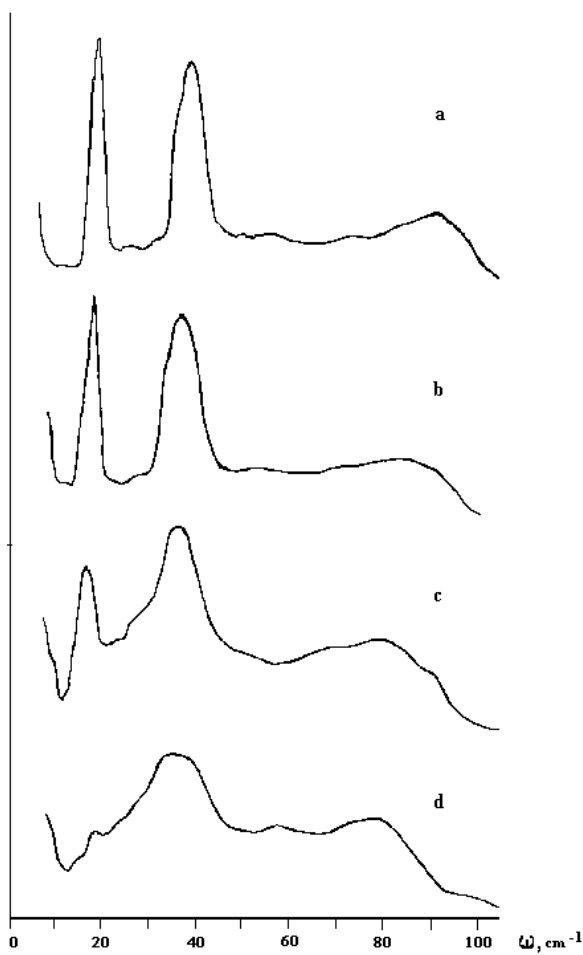

Fig.2. Low – frequency spectrums of powders of the investigated mix-crystals at a various content of components (a - 17.0%, b - 34.5%, c - 41.0% and d - 50.0% of a para-chloronitrobenzene)

Apparently, in range of concentrations of a para-chloronitrobenzene from 25.0% up to 50.0% the sum of spectrums of mixed crystal's rod frame consists of α and β phases is observed. At magnification of concentration of a para-chloronitrobenzene spectral intensity of a phase would monotonically decrease. It's explains a diminution of intensity of well observable line about 20.0 cm$^{-1}$ and gradual magnification of spectral intensity β phases, that it is visible about 90.0 cm$^{-1}$.

Thus, spectrums of low – frequencies of the mixed crystal of a para-dibromobenzene with a para-chloronitrobenzene in range of concentrations from 25.0% up to 50.0% of a para-chloronitrobenzene confirm presence simultaneously two α and β phases in this range that confirms presence of rod frame in a single crystal.

**References**

1. S.A. Remiga, P.M. Miasnikova, A.I. Kitaigorodskii, Crystallography **10**, p.875, 1965 (in Russian)
2. S.A. Remiga, P.M. Miasnikova, A.I. Kitaigorodskii, Journ. Struct. Chem. **10**, p.1131, 1969 (in Russian)
3. R.A. Laudise, The Growth of Single Crystals, Prentice-Hall Inc., New Jersey, 1970
4. K. Meyner, Physikalisch-chemische Kristallographie, VEB Deutscher Verlag fur Grundstoffindustrie, Leipzig, 1968